# Noninvasive nonlinear imaging through strongly-scattering turbid layers

Ori Katz,[1] Eran Small,[1] Yefeng Guan,[1,2] Yaron Silberberg,[1,*]

[1] Department of Physics of Complex Systems, Weizmann Institute of Science, Rehovot, Israel.

[2] State Key Laboratory of Optoelectronic Materials and Technologies, Sun Yat-sen University, Guangzhou 510275, China

*Corresponding author: yaron.silberberg@weizmann.ac.il

**Diffraction-limited imaging through complex scattering media is a long sought after goal with important applications in biomedical research. In recent years, high resolution wavefront-shaping has emerged as a powerful approach to generate a sharp focus through highly scattering, visually opaque samples. However, it requires a localized feedback signal from the target point of interest, which necessitates an invasive procedure in all-optical techniques. Here, we show that by exploiting optical nonlinearities, a diffraction-limited focus can be formed inside or through a complex sample, even when the feedback signal is not localized. We prove our approach theoretically and numerically, and experimentally demonstrate it with a two-photon fluorescence signal through highly scattering biological samples. We use the formed focus to perform two-photon microscopy through highly scattering, visually opaque layers.**

The inherent inhomogeneity of complex samples such as biological tissues induces light scattering, which limits the resolution of light focusing and poses a major hurdle to deep-tissue microscopy[1]. Adaptive optics techniques are very effective in correcting the aberrations induced by sample index mismatch and thin tissues[2-4]. Until recently these techniques were considered impractical for turbid and thick multiply scattering samples, where no unscattered 'ballistic' light components remain and the propagating light is diffused to form complex speckle patterns with no simple relation to the incident wavefront[5,6]. This conception changed after the 2007 work of Vellekoop and Mosk[7,8] where it was shown that high resolution wavefront shaping can be used to create a high intensity diffraction-limited focus deep in the diffusive light propagation regime, where essentially no ballistic photons are present, such as through a thick layer of white paint or the shell of an egg.

Diffraction-limited focusing by wavefront shaping is attained even when the number of degrees of control is much smaller than the number of scattered modes, i.e. with a correction that is far from being perfect[7-12]. Indeed, as a result, only a small fraction of the initial light energy is actually focused by this technique, but the high contrast focus can still be very valuable for many applications, and in particular for nonlinear microscopy. The main limitation of wavefront shaping techniques is that a direct feedback from the target point is required, either by directly observing it with a camera[7,14] or by embedding a 'guide star' or a detector at the target position[12-13]. Both these requirements are not compatible with most imaging applications. Very recently the technique was combined with acoustics to obtain the feedback signal non-invasively[15-20], though with focal dimensions exceeding the optical wavelength.

Here we show that nonlinear optical feedback, such as the one widely used for two-photon microscopy, can be used in epi-detection geometry to obtain diffraction-limited focusing through strongly scattering turbid samples. Furthermore, we demonstrate that this focus can be used for nonlinear imaging (here demonstrated in two-photon microscopy) in reflection-mode in a totally noninvasive manner. We theoretically and experimentally prove that one can focus an ultrashort pulse on an object hidden behind a scattering medium by optimizing the *total nonlinear signal* that is produced in the specimen. This technique should be adaptable to most nonlinear microscopy techniques, such as two-and three-photon fluorescence (2PF, 3PF), second- and third-harmonic generation (SHG, THG), four-wave-mixing and Coherent Anti-Stokes Raman Scattering (CARS). We show that the optimization results in a single, tightly-focused spot, even when the optimized signal is collected from a very large area on the object, specifically by epi-detection through the same scattering medium. Nonlinear optimization has been shown before to correct also for the temporal distortions of ultrashort pulses propagating through scattering media[21], hence the generated focus is also short in temporal duration and hence intense. Furthermore, by raster-scanning this focal spot, exploiting the so called memory effect[22-23], an image could be collected from the vicinity of the optimized focus[11-12,24]. Such a result cannot be obtained by optimizing a linear signal, such as total intensity of fluorescence. We believe that these results form an important step towards microscopic optical imaging deep inside scattering tissues.

**Experimental results**

Figure 1(a) shows the setup we have used for demonstrating our technique with two-photon excited fluorescence microscopy, one of the most widely used nonlinear imaging technique in biology and neuroscience[25]. 100 fs long pulses from a Ti:Sapphire laser illuminate the object after passing through a phase-only spatial light modulator. As a first experiment an optical diffuser was used to introduce scattering, and a thin two-photon homogeneous fluorescence screen served as the distributed object. The total excited fluorescence signal is epi-detected through the scattering medium with a single sensitive integrating detector. The spatial distribution of the two-photon excited fluorescence on the fluorescent screen target is inspected by a camera placed on the other side of the medium. Fig. 1(b) shows the speckely spatially-scattered fluorescence pattern recorded from the screen prior to optimization. We then used an optimization procedure based on a genetic algorithm[26] to find the SLM phase pattern that maximizes the total two-photon fluorescence signal measured in epi-detection. Fig. 1(c) shows the resulting optimized focus point on the screen. The progress of total detected signal during the optimization process is shown in Fig. 1(d) and in Supplementary Video 1. As can be observed in Fig.1c by optimizing the total backward-scattered 2PF signal collected via the illumination optics with a large-area detector, the excitation pulses focused to a single diffraction limited focal spot on the object, *though its location is not controlled or predetermined.* Note that the total enhancement in the signal is rather small – about 50% over its initial value, although the focused intensity is enhanced by a considerably larger factor.

Once a sharp spot is formed, it can be raster-scanned to obtain a microscopic image of an object hidden behind the scattering layer. The image collected point-by-point in a similar manner to image formation in standard two-photon microscopy, by scanning the focal point through the sample, for example by using the same SLM as a scanning device, although fast scanning mirrors systems can be utilized for increased speed. The field-of-view of this imaging technique is limited by the range where the SLM wavefront correction is effective, which is dictated in diffusive media by the 'optical memory-effect'[22-24], or the isoplanatic patch size[27]. The effective field-of-view will be especially large in cases when the random distorting medium is a thin layer located at a distance from the imaging target plane[24]. Note that for the scanning to be effective it is crucial to conjugate the correcting SLM plane with the distorting scattering layer plane[11-12,24],

An experimental demonstration of focusing and raster imaging using this technique is presented in Figure 2. In this experiment, the fluorescent screen is replaced by a cluster of 2-photon fluorescence Fluorescein crystallites on a glass cover-slip, placed behind the optical diffuser. After the same optimization procedure, the formed focus (whose exact position on the object is unknown and depends on the initial conditions and fluorescent structure), was scanned by adding linear phase ramps on top of the SLM correction phase pattern. The image that was obtained with the optimized focus (Fig. 2b) is not just of higher resolution and contrast, when compared to the blurred standard two-photon microscopy image, but is also approximately an order of magnitude brighter than an image obtained without optimization (Fig 2a). The true image of the object, with the diffuser removed, is shown in Fig.2c

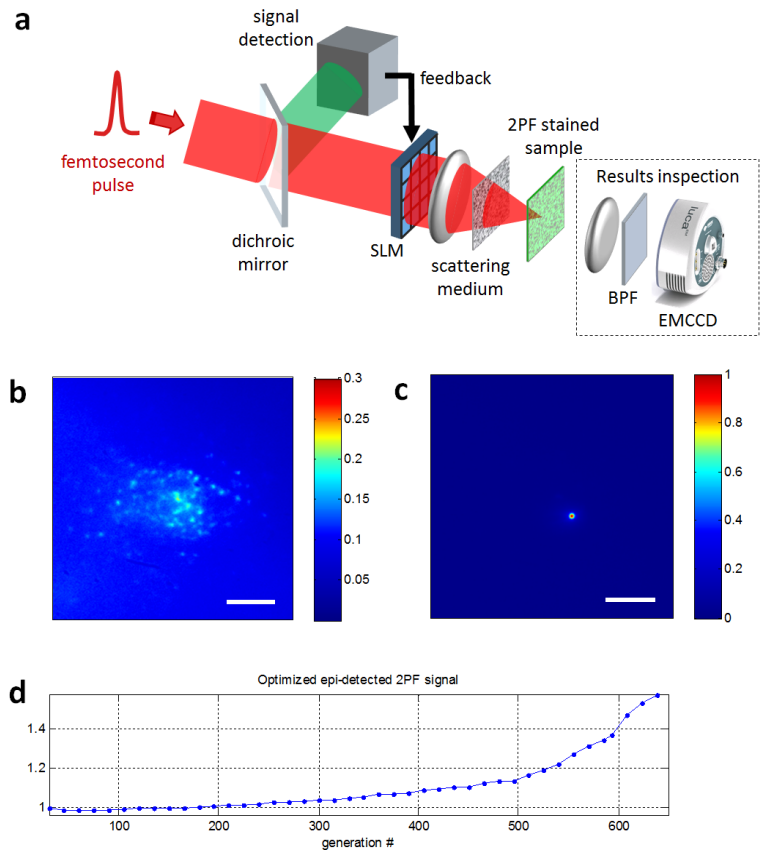

**Fig. 1**. Noninvasive focusing through scattering samples: (a) Experimental system: 100 fs pulses are sent via a spatial light modulator (SLM) and focused through a diffuser on a two-photon fluorescent screen. The fluorescence is collected via the same optics and used as a signal for an optimization algorithm. An auxiliary imaging system records the speckle image on the screen. (b) Speckle pattern recorded from the 2PF screen through an optical diffuser before optimization. (c) Optimized focus via maximization of total epi-detected 2PF. (d) Progress of the optimization process.

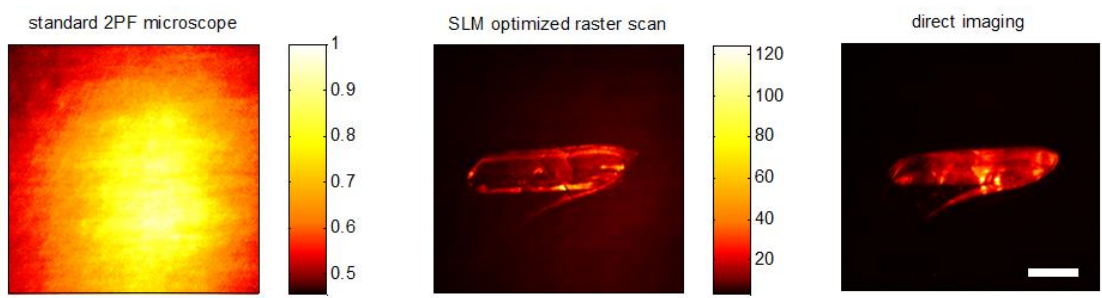

**Fig. 2.** Two-photon imaging through an optical diffuser using the memory-effect: (a) standard 2PF microscopy image of an object (cluster of Fluorescein crystallites) as observed through a diffuser. (b) after optimizing the total 2PF signal, a focus is formed. A bright and crisp image of the 2-photon object is obtained by collecting the signal in epi-geometry, while the focus is raster-scanned by adding linear phase-ramps to the SLM phase-pattern. (c) A transmission microscope image of same object without the scattering medium is presented for comparison. Scale bar 100μm.

**Discussion**

The experimental results show that optimizing a nonlinear signal is a promising path to forming a diffraction-limited spot on a planar object in the diffusive light propagation regime, where no ballistic unscattered light is present. This result is related to recent results in adaptive optics, where two-photon optimization is used to refocus light at depths of an order of one transport mean free path[28], where an initial focus is present in the unoptimized field, and to nonlinear photoacoustic focus formation on a planar target[29]. However, as we show below, careful analysis of the focus formation process shows that in the general case of a three-dimensional fluorescent object located in the diffusive regime *a higher order (N>2) nonlinearity is required to assure focusing to a single speckle grain.* For a linear signal, as in standard fluorescence imaging, energy conservation dictates that the total integrated signal would not change when the energy is focused tightly or spread over a large area, if the fluorophores are distributed homogeneously in the specimen. In contrast, an integrated *nonlinear* signal does not obey such a conservation law, and the total integrated signal could be larger the tighter the beam is focused, depending on the order of nonlinearity. Simple geometrical considerations can help to determine the range where this mechanism could be applied.

To obtain some insight, consider first a simple planar nonlinear object producing an *N*-th order nonlinear signal, and assume that the incoming beam is scattered to a number of speckles $N_{speckles}$ on the object. The total generated signal power would be proportional to:

$$P_{tot} \propto N_{speckles} \cdot \left(\frac{P_{laser}}{N_{speckles}}\right)^N = P_{laser}^N \left(\frac{1}{N_{speckles}}\right)^{N-1} \quad (1)$$

where $P_{laser}$ is the total beam power on the sample. Clearly, for a linear signal, i.e. *N*=1, there is no advantage in reducing the number of speckles since the total signal is independent of the size of the beam (i.e. $N_{speckles}$), as we have argued above (Eq. 1). However, the situation is remarkably different in the case of a nonlinear signal, i.e. *N*>1, where the total signal power would maximize for a beam where $N_{speckles}$ is minimized, i.e. for the most focused beam ($N_{speckles}$=1).

The fact that our samples were thin (either the uniform 2PF screen used in Fig. 1, or the sample used in Fig. 2) was important for this process to work. To understand the issue with a thick nonlinear medium, it is instructive to consider a Gaussian beam focused to a waist $w_0$ inside such a thick medium, producing an $N^{th}$-order nonlinear incoherent signal, such as N-photon fluorescence. In that case, for *N*>1 nonlinearity, the total nonlinear signal can be approximated by the signal generated inside a cylinder of volume *V* along the beam confocal parameter: $b=4\pi w_0^2/\lambda$. Under this assumption, the total generated (or detected) signal would be proportional to:

$$P_{tot} \propto V \cdot I_{laser}^N \propto \left(b \cdot \pi w_0^2\right) \cdot \left(\frac{P_{laser}}{w_0^2}\right)^N \propto w_0^4 \cdot \left(\frac{P_{laser}}{w_0^2}\right)^N \propto \frac{P_{laser}^N}{w_0^{2(N-2)}} \quad (2)$$

It is evident that for *N*=2, e.g. 2PF, the total signal is constant, independent of $w_0$. Hence optimization in a thick, homogenous second-order medium is not possible. Only for *N*>2, i.e. only for a nonlinearity which is higher than a second-order, the signal increases by focusing. This should be contrasted with the result of Eq. 1 for a planar nonlinear object, where any nonlinearity of *N*>1 was sufficient to ensure focusing down to a single speckle grain. We note that this fact is closely related to harmonic generation in bulk crystals and the determination of optimal focusing[30].

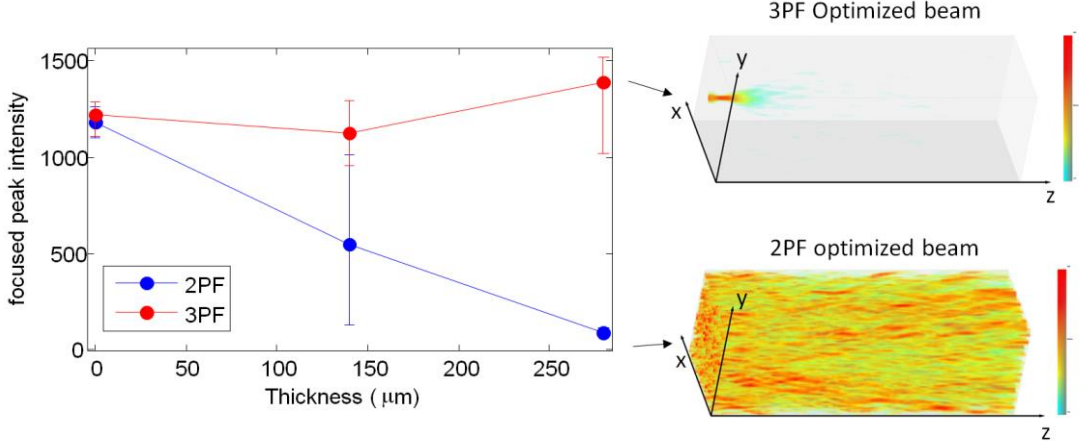

**Fig. 3.** Results of simulations of optimization of 2PF and 3PF signals. A speckle filed is simulated, and the calculated total nonlinear signal is then used as the feedback in a genetic algorithm for wavefront shaping. In (a) we show the resulting peak enhancement of such simulation with three different thicknesses of the nonlinear medium. In thin media both nonlinearities led to convergence and significant enhancement of the peak intensity. However, in thick media, only 3PF led to focusing and peak enhancement, as shown in (b). Two-photon effects did not converge and left the speckle filed practically unchanged.

To complement this naïve basic analysis, we performed exact numerical simulations that produce results which are on-par with this conclusion. Figure 3 shows results of simulations emulating the actual process: the speckle field was calculated by applying a random phase structure, and the nonlinear signal generated by the speckle field was integrated and used as feedback for optimization of wavefront correction using the same genetic algorithm. Note that in these simulations, as in our experiments, the field is first scattered by a thin layer and then propagated through the nonlinear medium, which was homogeneous and non-scattering. The simulations results, which are averaged over ten realizations of disorder for each case, show that 2PF does not lead to optimized focusing in thick three-dimensional objects, while 3PF worked well in all geometries.

Nonlinear feedback signal has been utilized previously for temporal compression; as has been shown in previous works[21,31,32], nonlinear optimization not only focuses the pulse in space, but also *compresses the pulse in time*, compensating for temporal distortions and resulting in a spatio-temporal focus. We expect therefore, that the optimized focal points used in this work are also temporally compensated, although we have not verified this by measurements.

Note that the analysis and simulations assumed a homogeneous distribution of the nonlinear medium, which is of course far from being realistic, and is actually representing a worst-case scenario. In actual samples, where the nonlinear agent is distributed unevenly in the specimen, it might well be that the algorithm would perform better, as concentrated region of high nonlinearity would serve as 'guide starts' and help the optimization process. As an example, we show in Figure 4 focusing through a 1 mm slice of brain tissue, on a nonlinear object placed behind it. The algorithm finds an optimum focus, which is, not surprisingly, located on a spot of highly concentrate fluorophore. Interestingly, we have observed that as the signal of one bright spot eventually fades due to photobleaching, the search algorithm will switch to another focal spot on another crystallite, bleaching it as well, and so on. We have confirmed in numerical simulations that two-photon nonlinearity is sufficient in many such cases where the volumetric nonlinear object is sparsely tagged or whenever there is even a weak ballistic unscattered component in the scattered beam, on par with the recent experimental results of Tang et al.[28].

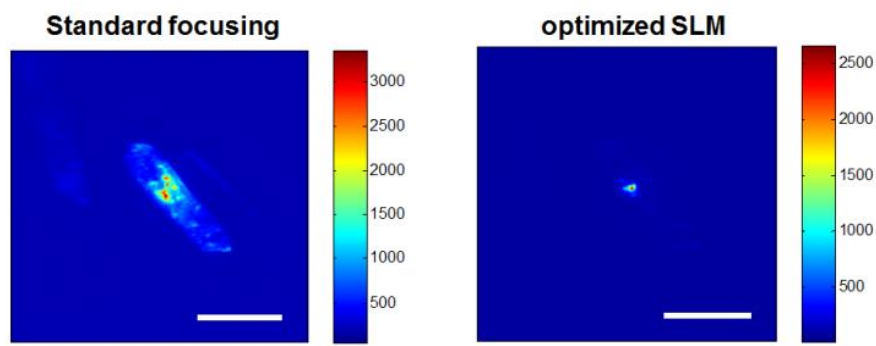

**Fig. 4.** Experimental focusing through a 1mm thick brain tissue. An object (Coumarin crystallites) is placed behind it, and optimization by maximizing the total two-photon fluorescence leads to a focal point which is actually localized on one of the strongest fluorescing crystallite.

The above analysis and discussion assumed incoherent nonlinear process, in particular two- and three-photon fluorescence. In coherent processes, such as second and third harmonic generation and CARS, phase matching consideration could have a significant effect, and we will not discuss them here, but generally we expect the above conclusions to be valid in cases where the phase-matching length is not large.

## Supplementary Information:

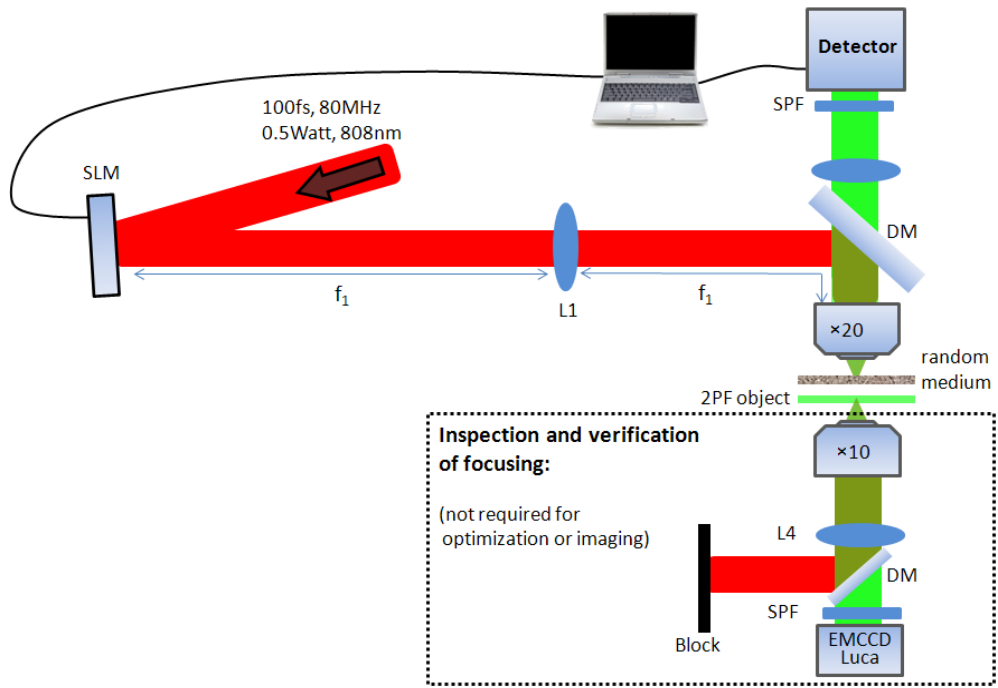

Fig. S1. Experimental setup for focusing through a scattering medium in epi-illumination and detection: the optimization is obtained by controlling the phase-only SLM to optimize the total 2PF signal collected by a wide-area detector utilizing a standard epi-detection two-photon microscope. The detector is implemented by integrating the total optical intensity measured by an Andor iXon EMCCD camera. The distance between the scatterer and the 2PF material was about 2 mm. In the proof-of-concept experiments, an additional imaging system (inspection and verification block in the figure), is used to prove that indeed the pulse has been focused on the target object. After the optimization, the focus can be scanned on the sample to obtain 2PF microscopy image of the object. Results are given in Fig. 2 of the main text.